\documentclass[a4paper,11pt]{article}
\usepackage{pos}
\usepackage{amsmath}
\usepackage{braket}
\usepackage{tikz-feynman}

\let\OLDthebibliography\thebibliography
\renewcommand\thebibliography[1]{
\OLDthebibliography{#1}
\setlength{\parskip}{0pt}
\setlength{\itemsep}{0pt plus 0.3ex}
}

\title{The mixing of two-pion and vector-meson states using staggered fermions}

\author*[a]{Fabian J. Frech}
\author[b,c]{Finn M. Stokes}
\author[a,c]{Kalman K. Szabo}
\author[a]{Balint C. Toth}
 \onbehalf{for the Budapest-Marseille-Wuppertal collaboration}

\affiliation[a]{University of Wuppertal,\\
Wuppertal, Germany}
\affiliation[b]{University of Adelaide,\\
Adelaide, Australia}
\affiliation[c]{FZ Juelich,\\
 Juelich, Germany}

\emailAdd{frech@uni-wuppertal.de}

\abstract{

In this study we employ staggered fermions to calculate the two-pion taste singlet states at rest. Leveraging the Clebsch-Gordan coefficients of the symmetry group associated with staggered fermions, we effectively compute the $\pi\pi$ contributions to the resting $\rho$-meson correlator. To discern the distinct energy states involved, we adopt a generalized eigenvalue problem-solving approach. This work will provide insight into the important role played by the two-pion contribution to the anomalous magnetic moment of the muon.

In this paper we present our group theoretic considerations and preliminary results on the contribution of two-pion states to the rho meson correlation function.
}

\FullConference{%
 40th International Symposium
on Lattice Field Theory\\
  July 31, 2023 to August 4, 2023\\
  Fermilab, Batavia, Illinois
}

\begin{document}
\maketitle

\section{Introduction}
In the pursuit of understanding the anomalous magnetic moment of the muon and its intriguing deviations from the predictions of the Standard Model, advanced theoretical frameworks and precise computational methods have become indispensable. Among these methodologies, lattice Quantum Chromodynamics (QCD) stands as a powerful tool that enables the exploration of the nonperturbative regime of QCD, particularly in understanding the hadronic contributions to $g-2$ \citep{Borsanyi:2020mff,Chao:2023lxw,Toth:2023eql}.

The persistent disparity between the measured and predicted values of has instigated meticulous investigations into the diverse contributions from the quantum vacuum. In this context, the two-pion exchange process, encapsulating intricate hadronic dynamics, emerges as a critical component that demands careful scrutiny. In general, pion exchange and interaction processes are an important part of current research \citep{Lahert:2021xxu,Dengler:2023szi,Grebe:2023tfx,Raposo:2023oru,Yu:2023xxf} as well as the hadronic light-by-light contributions to muon $g-2$ \citep{Zimmermann:2023tal,Gerardin:2023naa}.

In this study we will investigate the mathematical structure of the two-pion coupling to the vector current for staggered fermions in order to reconstruct the vector correlation function for comparatively large time separations.

In the following section we will give a short review on the symmetry group of staggered fermions and explain the mathematical formalism we used to construct the two-pion contributions to the $\rho$ correlation function. We explain our formulas and derivations and give the explicit values of the coefficients in the appendix. In the third section a short overview of the simulation setup and the implmentation of two-pion and vector meson correlators is given. Afterwards we show a few preliminary simulation results on the reconstruction of the $\rho$-correlator. In the fifth section we discuss our results and give an outlook to further investigations.

\section{Constrcution of vector states from moving pseudo-scalar states}
\label{sec:maths}
In this section, we aim to introduce the mathematical framework utilized for constructing $\pi\pi$ correlation functions whose quantum characteristics align with those of the $\rho$. To achieve this, an exploration of the symmetries inherent in the free staggered action is necessary. This action, denoted as $S_{st}$, is defined as follows:
\begin{align}
S_{st} = \sum_{x \in \Lambda}\bar \psi_x \left( m \psi_x +\sum_{\mu}\eta_{\mu}(x)\frac{\psi_{x+\hat\mu} - \psi_{x-\hat\mu}}{2}\right)\quad \textrm{with}\quad
 \eta_{\mu}(x) = (-)^{\sum_{\nu < \mu}x_{\nu}}.
\end{align}
These symmetries consist of transformations involving shifts $S_{\mu}$ by one lattice spacing in direction $\mu$, rotations $R_{\mu\nu}$ in the $\mu\nu$-plane by $\frac{\pi}{2}$, spatial inversion $I_S$, charge conjugation $C_0$, and taste transformations $\Xi_{\mu} = \frac{S_{\mu}}{\vert S_{\mu} \vert}$ \citep{Kilcup:1986dg,Golterman:1984dn}.

The symmetry group $\mathcal{G}$ is integral to our analysis. To measure correlation functions effectively, we focus our symmetry considerations on a fixed time slice. Consequently, we exclude the generators $S_4$ and $R_{i4}$. However, $\Xi_4$ remains a component of the reduced symmetry group $\mathcal{H}$. The irreducible representations (irreps) of $\mathcal{H}$ correspond to states possessing distinct quantum characteristics such as momentum, spin, parity, charge conjugation quantum number, and taste.

To examine the resting taste-singlet $\rho$-meson, we require the vector-representation of the three-dimensional Würfel group $W_3$ \citep{Baake:1981qe,Mandula:1983ut}, comprising rotations and spatial inversion (excluding charge conjugation since we aim for a $\pi\pi$ correlator with negative charge conjugation). The representation of a single pion with momentum $\vec p$ and taste $\xi_{\mu}$ is expressed as:
\begin{align}
D(\vec S,R,I_S,\Xi_{\mu})\ket{\vec p,\xi_{\mu}}_{\pi} = e^{i(-)^{I_S}R\vec p \cdot \vec S}(-)^{I_S}(-)^{R\vec \xi \cdot \vec \Xi + \xi_4\Xi_4}\ket{R\vec p ,R\vec \xi, \xi_4 }_{\pi},
\end{align}
Meanwhile, a two-pion correlation function is derived from the direct product of two single-pion correlators. Notably, this representation is reducible into subspaces determined by rotational orbits of taste and momentum \citep{Golterman:1986jf}. Completely reducible representations can be expressed as a direct sum over irreducible representations. Leveraging this, we precisely extract those two-pion correlators featuring the $\rho$-corrlator as one of their constituents:
\begin{align}
\ket{\vec p^1,\xi^1_{\mu}}_{\pi}\otimes \ket{\vec p^2,\xi^2_{\mu}}_{\pi} =\left( \bigoplus_{i = 1}^{a_{\rho}(\vec p^1,\xi^1_{\mu},\vec p^2,\xi^2_{\mu})}\ket{\vec p^1,\xi^1_{\mu},\vec p^2,\xi^2_{\mu}}_{\rho}\right) \oplus ...  
\end{align}
The coefficients $a_{\rho}$, represented by positive integers, are computable using the characters of the respective representations:\begin{align}
a_{\rho}(\vec p^1,\xi^1_{\mu},\vec p^2,\xi^2_{\mu}) = \frac{1}{\vert \mathcal{H}\vert}\sum_{h \in \mathcal{H}}\chi^*_{\rho}(h) \chi_{\pi\pi}(\vec p^1,\xi^1_{\mu},\vec p^2,\xi^2_{\mu};h)
\end{align}
Through algebraic computations, one can determine that these coefficients are solely non-zero if $\vec p^1 = -\vec p^2 \neq \vec 0$ and $\vec \xi^1 = \vec\xi^2$. The specific values are provided in Table \ref{tab:my_label}.
\begin{table}[]
    \centering
    \begin{tabular}{c||c|c|c|c|c}
                           &$\Vert \vec p \Vert^2 = 0$&$\Vert \vec p \Vert^2 = 1$&$\Vert \vec p \Vert^2 = 2$& $\Vert \vec p \Vert^2 = 3$& $\Vert \vec p \Vert^2 = 4$    \\
                           \hline\hline
         $\Vert \vec\xi \Vert^2 = 0$& 0&1&1&1&1\\
         $\Vert \vec\xi \Vert^2 = 1$& 0&2&3&2&2\\
         $\Vert \vec\xi \Vert^2 = 2$& 0&2&3&2&2\\
         $\Vert \vec\xi \Vert^2 = 3$& 0&1&1&1&1\\
    \end{tabular}
    \caption{Multiplicities of the vector irreps in the respective $\pi\pi$-states. The numbers are independent of $\xi_4$.}
        \label{tab:my_label}
\end{table}
The overlap between different correlation functions can be expressed using the Clebsch-Gordon coefficients \citep{Lahert:2021xxu}:
\begin{align}
\ket{\{\vec p\},\{\xi_{\mu}\}}_{\rho^\alpha} = \sum_{\vec p \in \{\vec p\}}\sum_{\vec \xi \in \{\vec \xi\}} C^\alpha(\vec{p},\vec{\xi}) \ket{\vec p, \vec \xi, \xi_4}_{\pi}\otimes \ket{-\vec p, \vec \xi, \xi_4}_{\pi} + ...,
\end{align}
Here, $\{\vec x\}$ denotes the orbit of $\vec x$ under lattice rotations and reflections. Utilizing mathematical techniques applicable to finite groups \citep{Sakata:1974hd,Dudek:2012gj}, we can compute the Clebsch-Gordon coefficients for this context. The specific values for these coefficients are detailed in appendix \ref{CG}.

\section{Implementation of the $\pi\pi$-states}
We utilized 48 Symanzik improved gauge configurations employing $2+1+1$ 4\texttt{stout} one-link fermions for our tests \citep{PhysRevD.69.054501}. The simulations were conducted on $32^2\times 64$ lattices, utilizing a gauge coupling parameter of $\beta = 3.7000$. This particular parameter corresponds to a lattice spacing of $0.1315\,$fm. Our chosen quark masses are positioned around the physical point ($m_l = 0.00205$, $m_s = 0.05729$, and $m_c = 0.67890$) \citep{Borsanyi:2020mff}.

We focused on measuring the propagator of the two-pion correlator, projected to negative charge conjugacy, specifically $\frac{\pi^+(x)\pi^-(y) - \pi^-(x)\pi^+(y)}{2}$. This $\pi^+\pi^-$ combination results in three distinct diagrams for the two-pion correlation function: the connected, the free, and the disconnected (see figure \ref{fig:diagrams}). Fortunately, due to the anti-symmetrization, the disconnected diagram vanishes identically.
\begin{figure}[h]
\begin{center}
\begin{tikzpicture}[very thick,q0/.style={->,DarkBlue,semithick,yshift=5pt,shorten >=5pt,shorten <=5pt}]

  \def\radius{0.9}
 
  \draw[] (0,-\radius) -- (2*\radius,-\radius) ;
  \draw[] (0,-\radius) -- (0,\radius) ;
  \draw[] (2*\radius,-\radius) -- (2*\radius,\radius) ;
  \draw[] (0,\radius) -- (2*\radius,\radius) ;
    \node[left] (1) at (0,0) {$u$};
        \node[right] (1) at (2*\radius,0) {$u$};
    \node[above] (1) at (\radius,\radius) {$\bar d$};
    \node[below] (1) at (\radius,-\radius) {$\bar d$};
    \node[above] (1) at (-0.2*\radius,\radius) {$\pi^{\dagger}(\vec x,0)$};
    \node[below] (1) at (-0.2*\radius,-\radius) {$\pi(\vec y,0)$};
    \node[above] (1) at (2.2*\radius,\radius) {$\pi(\vec z,t)$};
    \node[below] (1) at (2.2*\radius,-\radius) {$\pi^{\dagger}(\vec w,t)$};
    \node[below] (1) at (\radius,-1.8*\radius) {$a)$};
\def \offset{3.5}
\draw[] (\offset,-\radius) -- (\offset+2*\radius,-\radius) ;
  \draw[] (\offset,-0.9*\radius) -- (\offset+2*\radius,-0.9*\radius) ;
  \draw[] (\offset,\radius) -- (\offset+2*\radius,\radius) ;
  \draw[] (\offset,0.9*\radius) -- (\offset+2*\radius,0.9*\radius) ;
    \node[below] (1) at (\offset+\radius,\radius) {$u$};
    \node[above] (1) at (\offset+\radius,\radius) {$\bar d$};
    \node[below] (1) at (\offset+\radius,-\radius) {$u$};
    \node[above] (1) at (\offset+\radius,-\radius) {$\bar d$};
    \node[above] (1) at (\offset-0.2*\radius,\radius) {$\pi^{\dagger}(\vec x,0)$};
    \node[below] (1) at (\offset-0.2*\radius,-\radius) {$\pi(\vec y,0)$};
    \node[above] (1) at (\offset+2.2*\radius,\radius) {$\pi(\vec z,t)$};
    \node[below] (1) at (\offset+2.2*\radius,-\radius) {$\pi^{\dagger}(\vec w,t)$};
        \node[below] (1) at (\offset + \radius,-1.8*\radius) {$b)$};

\begin{feynman}
    \vertex (a) at (\offset+0.6*\radius,\radius);
    \vertex (b) at (\offset+1.7*\radius,-\radius);
    \diagram*{
      (a) -- [gluon] (b);
    };
\end{feynman}

    \draw[] (2*\offset,-\radius) -- (2*\offset,+\radius) ;
  \draw[] (2*\offset + 0.1*\radius,-\radius) -- (2*\offset + 0.1*\radius, +\radius) ;
 \draw[] (2*\offset+2*\radius,-\radius) -- (2*\offset+2*\radius,+\radius) ;
  \draw[] (2*\offset + 1.9*\radius,-\radius) -- (2*\offset + 1.9*\radius, +\radius) ;
    \node[left] (1) at (2*\offset,0) {$u$};
    \node[right] (1) at (2*\offset,0) {$\bar d$};
    \node[left] (1) at (2*\offset+2*\radius,0) {$u$};
    \node[right] (1) at (2*\offset+2*\radius,-0) {$\bar d$};
 \begin{feynman}
    \vertex (a) at (2*\offset,0.6*\radius);
    \vertex (b) at (2*\offset+2*\radius,-0.6*\radius);
    \diagram*{
      (a) -- [gluon] (b);
    };
    \end{feynman}
    
    \node[above] (1) at (2*\offset-0.2*\radius,\radius) {$\pi^{\dagger}(\vec x,0)$};
    \node[below] (1) at (2*\offset-0.2*\radius,-\radius) {$\pi(\vec y,0)$};
    \node[above] (1) at (2*\offset+2.2*\radius,\radius) {$\pi(\vec z,t)$};
    \node[below] (1) at (2*\offset+2.2*\radius,-\radius) {$\pi^{\dagger}(\vec w,t)$};
            \node[below] (1) at (2*\offset + \radius,-1.8*\radius) {$c)$};

\end{tikzpicture}

\begin{tikzpicture}[very thick,q0/.style={->,DarkBlue,semithick,yshift=5pt,shorten >=5pt,shorten <=5pt}]

  \def\radius{0.9}
 
  \draw[] (0,0) -- (2*\radius,-\radius) ;
  \draw[] (2*\radius,-\radius) -- (2*\radius,\radius) ;
  \draw[] (0,0) -- (2*\radius,\radius) ;
    \node[right] (1) at (2*\radius,0) {$\bar d$};
    \node[above] (1) at (\radius,0.5*\radius) {$u$};
    \node[below] (1) at (\radius,-0.5*\radius) {$u$};
    \node[left] (1) at (0,0) {$\rho^{\dagger}(\vec x,0)$};
    \node[above] (1) at (2.2*\radius,\radius) {$\pi(\vec z,t)$};
    \node[below] (1) at (2.2*\radius,-\radius) {$\pi^{\dagger}(\vec w,t)$};
            \node[below] (1) at (\radius,-1.8*\radius) {$d)$};
   \def\offset{5*\radius}
 \draw[] (\offset,0) -- (\offset + 2*\radius,0) ;
  \draw[] (\offset,0.1) -- (\offset + 2*\radius,0.1) ;
  \draw[] (0,0) -- (2*\radius,\radius) ;
    \node[above] (1) at (\offset +\radius,0.2*\radius) {$u$};
    \node[below] (1) at (\offset +\radius,-0.2*\radius) {$\bar d$};
    \node[left] (1) at (\offset +0,0) {$\rho^{\dagger}(\vec x,0)$};
    \node[left] (1) at (\offset +3.5*\radius,0) {$\rho(\vec z,0)$};
        \node[below] (1) at (\offset + \radius,-1.8*\radius) {$e)$};

\end{tikzpicture}
\end{center}
\caption{The diagrams corresponding to $\pi\pi/\rho$ propagators.  The first one (conn., $a)$) includes interchange of quarks in the two pions, the second one (free, $b)$) describes two pions moving separately and the third one (disc., $c)$) vanishes identically due to anti-symmetry in the pions. The triangle ($d)$) describes the decay of the vector mesons into two pions and the last diagram ($e)$) is the vector meson correlation functions.}
\label{fig:diagrams}
\end{figure}
The correlation functions are measured by applying the inverse Dirac operator and taste (spin/momentum) operator on random Wall-sources. To incorporate momenta, complex phases are multiplied at each lattice point, followed by anti-symmetrization in the ingoing and outgoing momenta after measurement. The spin and taste structures adhere to the methodology outlined in \citep{Follana:2006rc}.

The free diagram is given by the direct product of two single-pion correlators. It's crucial to note that while the ingoing momentum or taste may differ from the outgoing ones within individual states, the total momentum and taste must be conserved overall.

The connected part is relatively more resource-intensive as it necessitates an inversion on every time-slice.

\section{Measurements and the GEVP}
In our investigation, we consider not only the vector meson but also all two-pion states possessing lower energy. These states correspond to orbits characterized by $\vert \vec p \vert = 1$ and $\xi_4 = 1$, alongside either $\vert \vec \xi \vert = 3$ or $\vert \vec \xi \vert = 2$. These specific correlators are depicted in the first line of Table \ref{tab:CG}. Hereafter, we refer to them as the pseudo-scalar and the parallel or perpendicular pseudo-vector, depending on the relative alignment of taste and momentum. By utilizing the diagrams illustrated in Figure \ref{fig:diagrams}, we proceed to construct the correlation matrix
\begin{align}
C(t) = \begin{pmatrix}
\rho(t)\rho^{\dagger}(0)& \rho(t+1)\rho^{\dagger}(0)& \pi\pi(t)\rho^{\dagger}(0)&...\\
\rho(t)\rho^{\dagger}(-1)& \rho(t+1)\rho^{\dagger}(-1)& \pi\pi(t)\rho^{\dagger}(-1)&...\\
\rho(t)\pi\pi^{\dagger}(0)& \rho(t+1)\pi\pi^{\dagger}(0)& \pi\pi(t)\pi\pi^{\dagger}(0)&...\\
...&...&...&...\\
\end{pmatrix}
\label{eq:correlation}
\end{align}
\begin{figure}
\centering
\includegraphics[width = 0.45\textwidth]{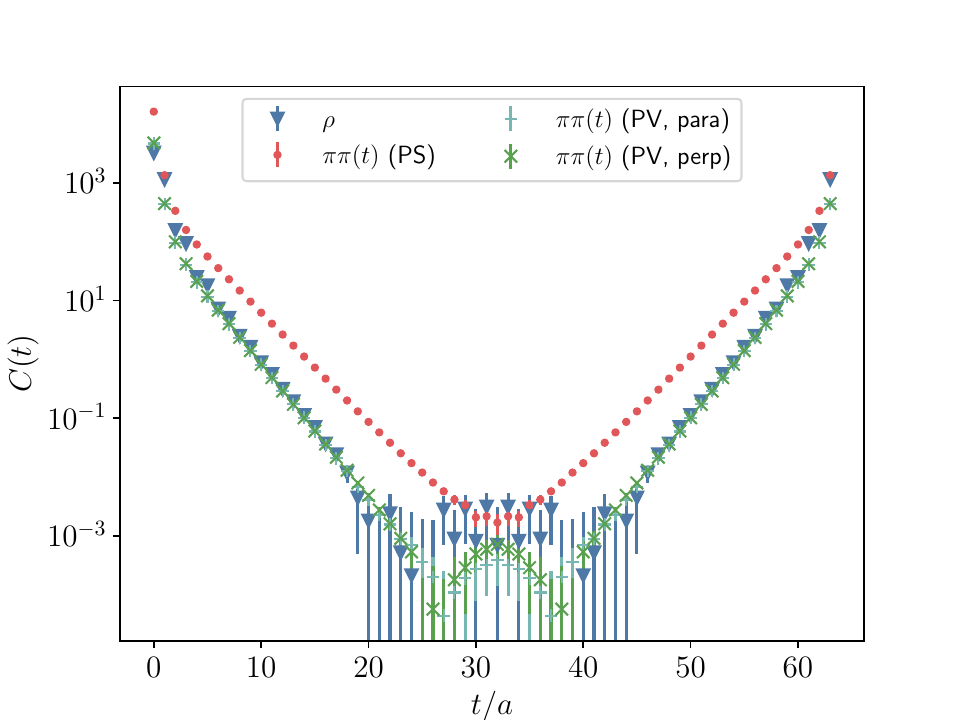}
\includegraphics[width = 0.45\textwidth]{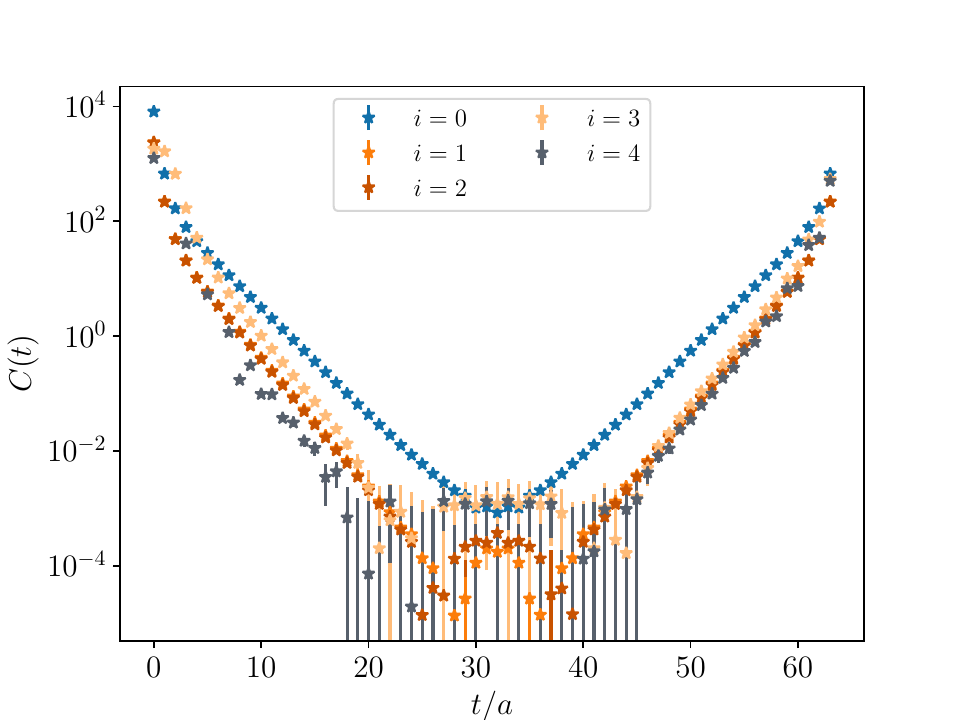}
\caption{Left hand side: The diagonal elements of the correlation matrix in Equation \ref{eq:correlation}.  The pion correlators are normalized by a factor of 0.01 at the sink and the source for a better visibility in the plot and a smaller condition number of the correlation matrix. This normalization will not affect the results. Right hand side: The eigenmodes of the correlation matrix.}
\label{fig:correlators}
\end{figure}
The time-shifted vector meson correlators have been incorporated to extract the oscillating parity partner using the pencil of functions method \citep{DeTar:2014gla}. The diagonal elements of the correlation matrix are illustrated in Figure \ref{fig:correlators}. To determine the eigenmodes of the system, we express it in terms of a Generalized Eigenvalue Problem (GEVP) \citep{Michael:1985ne,Luscher:1990ck}
\begin{align}
C(t_0 + dt)\cdot v_i = \lambda_i(t_0,dt)C(t_0)\cdot v_i
\end{align}
For this study, we set $t_0 = 2$ and $dt = 1$, computing a set of five eigenvectors $v_i$. The eigenmodes of the correlation matrix are subsequently computed using:\begin{align}
C_i(t) = v_i^T \cdot C(t) \cdot v_i
\end{align}
The five distinct eigenmodes are depicted on the right-hand side of Figure \ref{fig:correlators}. The increase in the number of eigenmodes is due to the inclusion of an additional parity partner oscillation through the pencil of functions method \citep{DeTar:2014gla}.

\begin{figure}
\centering
\includegraphics[width = 0.45\textwidth]{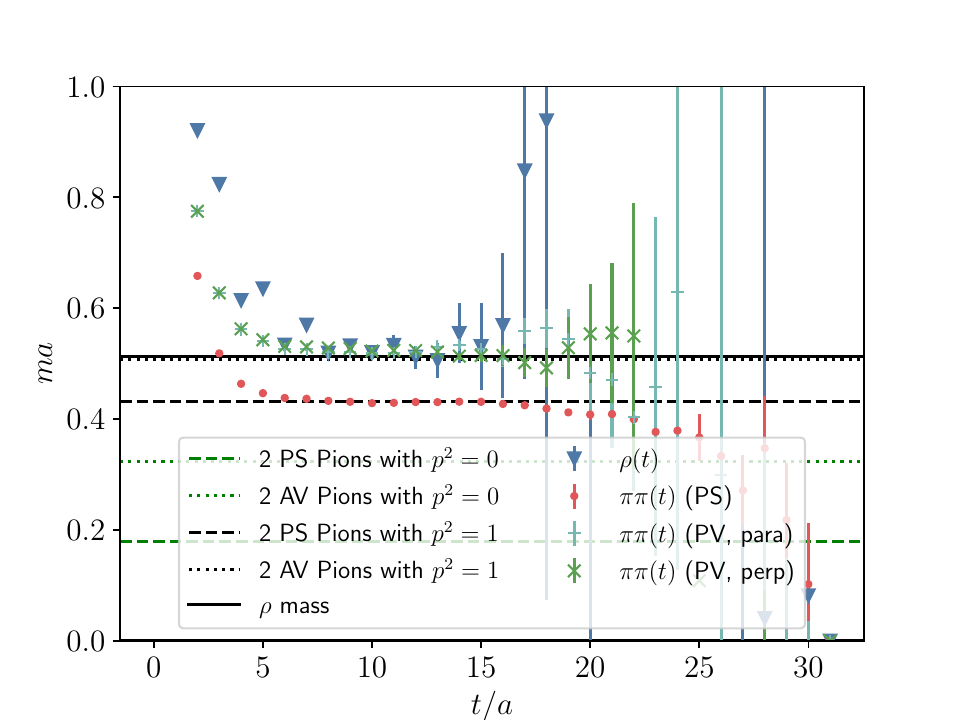}
\includegraphics[width = 0.45\textwidth]{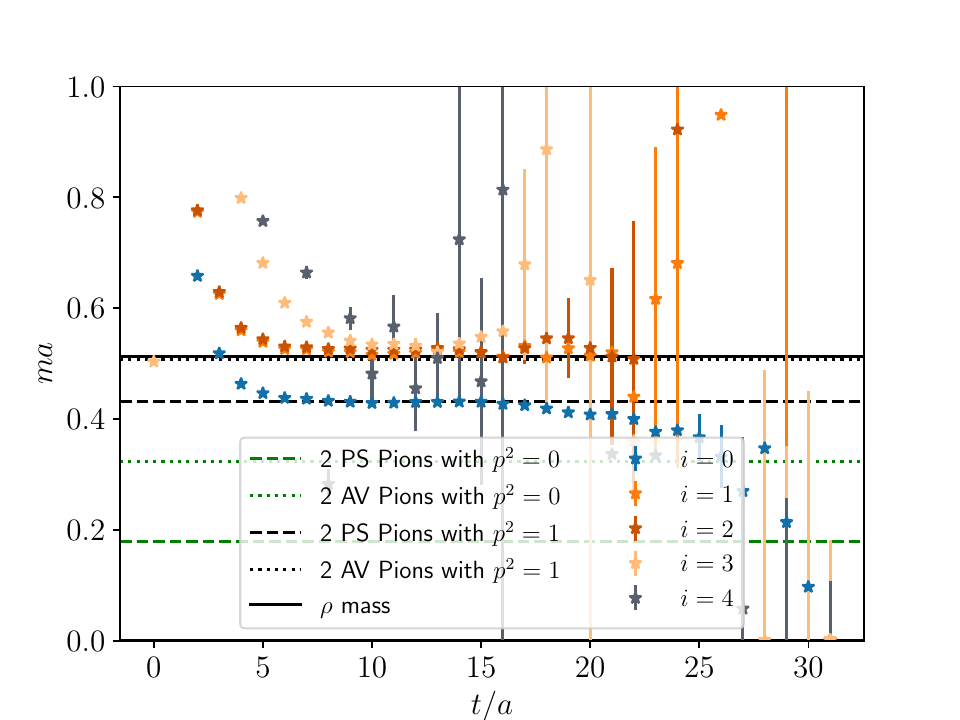}
\caption{Left hand side: The two-point effective mass of diagonal elements of the correlation matrix in Equation \ref{eq:correlation}. Right hand side: The two-point effective mass of the eigenmodes of the correlation matrix.}
\label{fig:mass}
\end{figure}

\begin{figure}
\centering
\includegraphics[width = 0.45\textwidth]{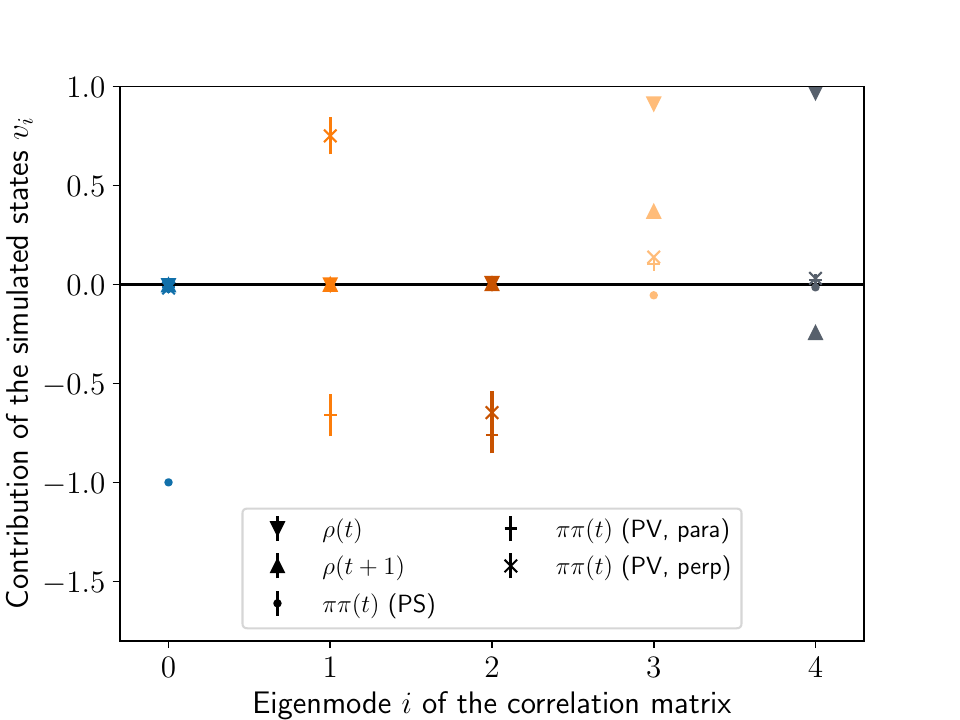}
\includegraphics[width = 0.45\textwidth]{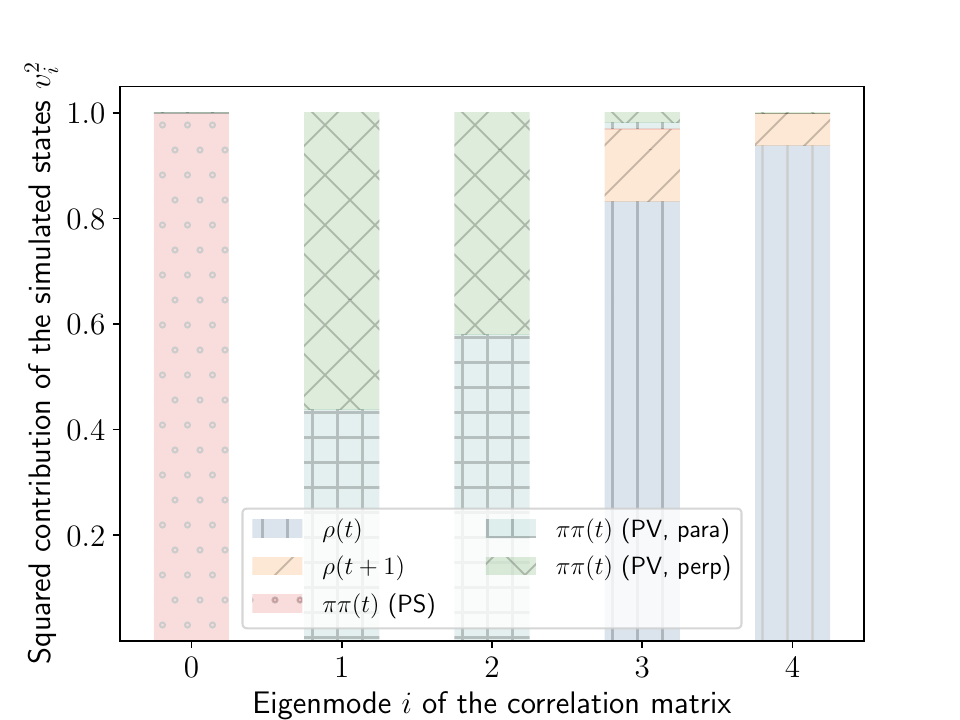}
\caption{Left-hand side: The eigenvectors of the GEVP. One can see the contribution of the input states to the eigenmodes. Right-hand side: The squared eigenvectors as a stacked bar plot.}
\label{fig:EV}
\end{figure}

Eigenmode $4$ represents the parity partner oscillation of the vector meson state. In Figure \ref{fig:mass}'s right-hand side, it is demonstrated that this mode does not form a definitive mass plateau. Moreover, as shown in Figure \ref{fig:EV}, both the shifted and unshifted vector mesons contribute with differing signs to this state.

State $3$ represents the pure vector meson state. While contributions from the shifted and unshifted vector meson correlators exist, there are additional minor contributions from the various two-pion correlators. The right-hand side of Figure \ref{fig:mass} illustrates that the effective mass of this state forms a plateau near the physical mass of the vector meson.

\begin{figure}
\centering
\includegraphics[width = 0.45\textwidth]{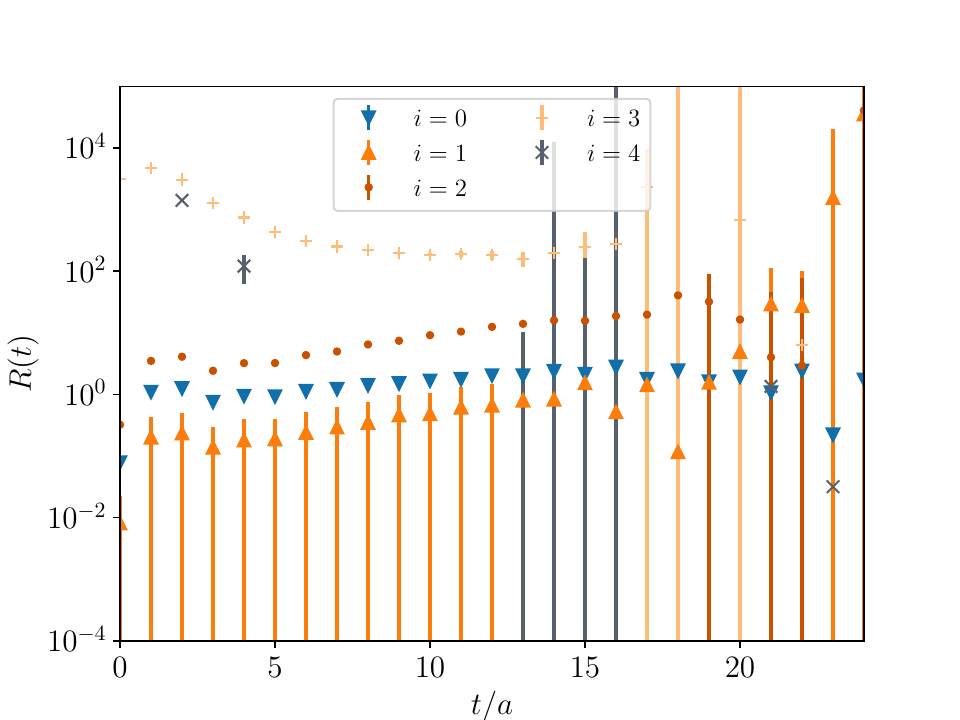}
\includegraphics[width = 0.45\textwidth]{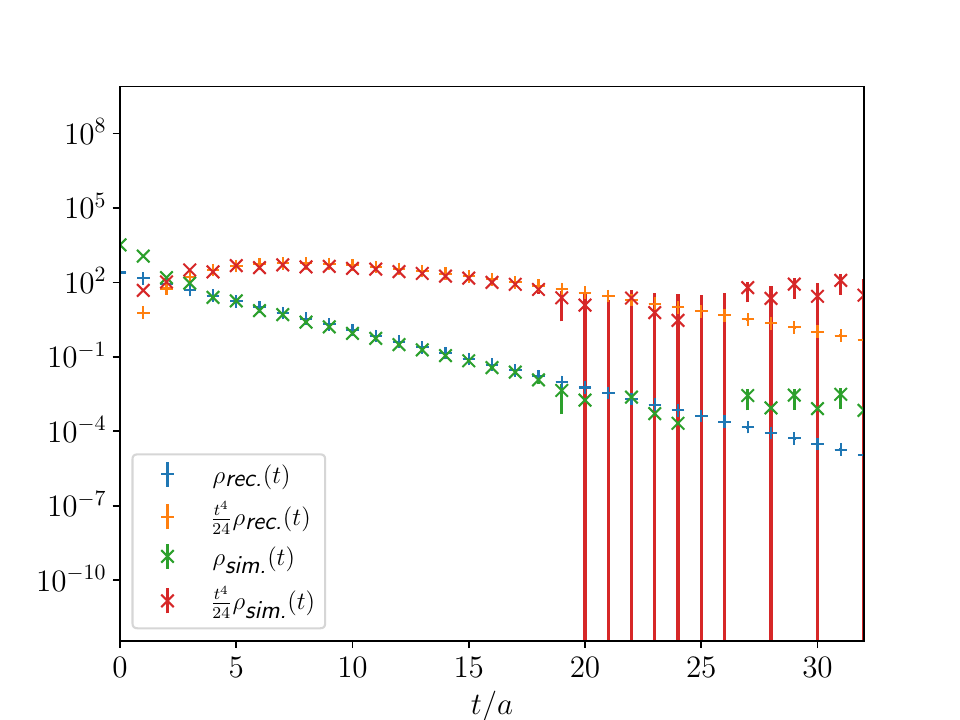}
\caption{Left-hand side: The effective coefficients of the reconstruction, the $4th$ will not be used in the reconstruction. Right-hand side: The reconstructed (\textit{rec.}) and the simulated (\textit{sim.}) vector correlators and the integrand of their first moment, which is the interesting observable for the computation of $g-2$.}
\label{fig:reconstruction}
\end{figure}

The pseudo-scalar state ($0$) remains relatively unaffected by the GEVP and exhibits a clear plateau in the effective mass close to its physical mass, determined by $2\sqrt{m_\pi^2 + \left(\frac{2\pi}{L} \right)^2 }$. In contrast, for the pseudo-vector states, their mass is determined by $2\sqrt{m_\pi^2 +\Delta_{TS}+ \left(\frac{2\pi}{L} \right)^2 }$, where $\Delta_{TS} = 40678\,\textrm{MeV}^2$ signifies the taste splitting observed in this ensemble. The two pseudo-vector states ($1$ and $2$) exhibit a degree of mixing, evident from Figure \ref{fig:EV}. However, as observed from Figure \ref{fig:mass}, both states showcase their mass plateau close to this physical estimator.\\
Our aim also involves reconstructing the vector meson correlator to analyze its long-range behavior, achieved by forming a linear combination of exponentially decaying functions:
\begin{align}
\rho_{rec.}(t) = \sum_{i = 0}^4 R_i \exp \left(-m_it\right)
\end{align} 
The coefficients $R_i$ are determined by plateaus fitted to:
\begin{align}
R_{\textit{eff},i} = \frac{\left(v_i\cdot C_{\rho}(t)\right)^2}{v_i^T \cdot C \cdot v_i \exp\left(-m_it \right)}
\end{align}
Here, $C_{\rho}$ represents the $\rho$-column of the correlation matrix. These coefficients are visualized on the left-hand side of Figure \ref{fig:reconstruction}. Notably, the parity partner state changes sign across different time-slices and is omitted from the final result, hence $R_4 = 0$. Additionally, one of the states of the axial pseudo-vectors aligns with zero. For state $3$, plateaus are fitted within the range $t/a \in [7,12]$, while for the remaining states $0-2$, the fitting range is $t/a \in [16,20]$.

The right-hand side of Figure \ref{fig:reconstruction} illustrates the total reconstruction and the first moment of the vector meson correlator.

\section{Conclusion and outlook}
This paper presents a derived mathematical framework for constructing $\pi\pi$ correlation functions that exhibit the quantum characteristics of a resting taste-singlet vector meson. The method was successfully tested using $48$ gauge configurations within a $\sim 4\,\textrm{fm}$ box, featuring a lattice spacing of $0.1315\,\textrm{fm}$ near the physical point. We managed to extract various contributing energy states below the vector meson mass and effectively reproduced the vector meson correlator for larger times.

Our next steps involve the reduction of simulation costs, particularly for the connected part, and subsequently enhancing statistics to attain more precise results. This improvement aims to refine the long-time contribution to g-2.\\

\textbf{Acknowledgements: }The computations were performed on HAWK at the High Performance Computing Center in Stuttgart and IRENE at Commissariat à l’Energie Atomique et aux énergies alternatives (CEA), France. We thank the Gauss Centre for Supercomputing, PRACE and GENCI (grant 52275) for awarding us computer time on these machines.
\appendix
\section{Clebsch-Gordan coefficients}
\label{CG}
In this section, we will present the Clebsch-Gordan coefficients for all taste orbits and momentum orbits up to $\vec p^2 = 4$. It's worth noting that there is no distinction between $\vec p^2 = 1$ and $\vec p^2 = 4$. Here, $e_i$ represents unit vectors in momentum space, while $f_i$ represents unit vectors in taste space, specifically when $\vert \vec p \vert = 1$ or $\vert \vec \xi \vert = 1$. For orbits with $\vert \vec \xi \vert = 2$, $f_i$ denotes a vector with a zero in the $i$th component and a one in the remaining components. $\vec 0$ signifies the zero vector when $\vert\vec \xi \vert = 0$ and a vector consisting only of ones when $\vert \vec \xi \vert = 3$. Moreover, $\lambda$, $\mu$, and $\nu$ can take the values $+1$ or $-1$. The Clebsch-Gordan coefficients are displayed in Table \ref{tab:CG}.

\begin{table}
\centering
\begin{tabular}{|c||c|c|}
\hline
&$\vec \xi^2 = 0(3)$&$\vec \xi^2 = 1(2)$\\
\hline\hline
$\vec p^2 = 1(4)$&$C^\alpha(\lambda \vec e_i,\vec 0) = \lambda \frac{1}{\sqrt{2}}\delta^{\alpha i}$&$C^\alpha(\lambda \vec e_i,\vec f_j) = \lambda \frac{1}{\sqrt{2}} \delta^{ij}\delta^{\alpha i}$\\
&&$C^\alpha(\lambda \vec e_i,\vec f_j) =  \frac{\lambda}{2}(1-\delta_{ij})\delta^{\alpha i}$\\
\hline
$\vec p^2 = 2$&$C^\alpha(\lambda \vec e_i + \mu e_j,\vec 0) =  \frac{\lambda}{2\sqrt{2}}\delta^{\alpha i}$&$C^\alpha(\lambda \vec e_i + \mu \vec e_j,\vec f_k) =  \frac{\lambda}{2\sqrt{2}}\vert\epsilon^{ijk}\vert\delta^{\alpha i}$ \\
&&$C^\alpha(\lambda \vec e_i + \mu \vec e_j,\vec f_k) = \frac{\lambda}{2\sqrt{2}}\delta^{ki}\delta^{\alpha i}$ \\
&&$C^\alpha(\lambda \vec e_i + \mu \vec e_j,\vec f_k) = \frac{\lambda}{2\sqrt{2}}\delta^{kj}\delta^{\alpha i}$ \\
\hline
$\vec p^2 = 3$&$C^\alpha(\lambda \vec e_i + \mu \vec e_j + \nu \vec e_k,\vec 0) =  \frac{\lambda}{2\sqrt{2}}\delta^{\alpha i}$&$C^\alpha(\lambda \vec e_i + \mu \vec e_j + \nu \vec e_k,\vec f_l) =  \frac{\lambda}{2\sqrt{2}}\delta^{il}\delta^{\alpha i}$ \\
&&$C^\alpha(\lambda \vec e_i + \mu \vec e_j + \nu \vec e_k,\vec f_l) =  \frac{\lambda}{4}\vert\epsilon^{jkl}\vert\delta^{\alpha i}$ \\
\hline
\end{tabular}
\caption{The Clebsch-Gordan coefficients needed for construcing the vector meson out of two-pion states.}
\label{tab:CG}
\end{table}

\bibliographystyle{JHEP}
\bibliography{proceedings.bib}

\providecommand{\href}[2]{#2}\begingroup\raggedright\begin{thebibliography}{10}

\bibitem{Borsanyi:2020mff}
S.~Borsanyi et~al.,
  \href{https://doi.org/10.1038/s41586-021-03418-1}{\emph{Nature} {\bfseries
  593} (2021) 51} [\href{https://arxiv.org/abs/2002.12347}{{\ttfamily
  2002.12347}}].

\bibitem{Chao:2023lxw}
E.-H. Chao, H.~B. Meyer and J.~Parrino,
  \href{https://arxiv.org/abs/2310.20556}{{\ttfamily 2310.20556}}.

\bibitem{Toth:2023eql}
B.~C. Toth, \href{https://doi.org/10.1051/epjconf/202328901005}{\emph{EPJ Web
  Conf.} {\bfseries 289} (2023) 01005}.

\bibitem{Lahert:2021xxu}
S.~Lahert, C.~DeTar, A.~X. El-Khadra et~al.,
  \href{https://doi.org/10.22323/1.396.0526}{\emph{PoS} {\bfseries LATTICE2021}
  (2022) 526} [\href{https://arxiv.org/abs/2112.11647}{{\ttfamily
  2112.11647}}].

\bibitem{Dengler:2023szi}
Y.~Dengler, A.~Maas and F.~Zierler,
  \href{https://arxiv.org/abs/2311.18549}{{\ttfamily 2311.18549}}.

\bibitem{Grebe:2023tfx}
A.~V. Grebe and M.~Wagman,  \href{https://arxiv.org/abs/2312.00321}{{\ttfamily
  2312.00321}}.

\bibitem{Raposo:2023oru}
A.~B.~a. Raposo and M.~T. Hansen,
  \href{https://arxiv.org/abs/2311.18793}{{\ttfamily 2311.18793}}.

\bibitem{Yu:2023xxf}
K.~Yu, Y.~Li, J.-J. Wu et~al.,
  \href{https://arxiv.org/abs/2311.03903}{{\ttfamily 2311.03903}}.

\bibitem{Zimmermann:2023tal}
C.~Zimmermann and A.~G\'erardin,
  \href{https://arxiv.org/abs/2311.10628}{{\ttfamily 2311.10628}}.

\bibitem{Gerardin:2023naa}
A.~G\'erardin, W.~E.~A. Verplanke, G.~Wang et~al.,
  \href{https://arxiv.org/abs/2305.04570}{{\ttfamily 2305.04570}}.

\bibitem{Kilcup:1986dg}
G.~W. Kilcup and S.~R. Sharpe,
  \href{https://doi.org/10.1016/0550-3213(87)90285-9}{\emph{Nucl. Phys. B}
  {\bfseries 283} (1987) 493}.

\bibitem{Golterman:1984dn}
M.~F.~L. Golterman and J.~Smit,
  \href{https://doi.org/10.1016/0550-3213(85)90138-5}{\emph{Nucl. Phys. B}
  {\bfseries 255} (1985) 328}.

\bibitem{Baake:1981qe}
M.~Baake, B.~Gemunden and R.~Odingen,
  \href{https://doi.org/10.1063/1.525461}{\emph{J. Math. Phys.} {\bfseries 23}
  (1982) 944}.

\bibitem{Mandula:1983ut}
J.~E. Mandula, G.~Zweig and J.~Govaerts,
  \href{https://doi.org/10.1016/0550-3213(83)90399-1}{\emph{Nucl. Phys. B}
  {\bfseries 228} (1983) 91}.

\bibitem{Golterman:1986jf}
M.~F.~L. Golterman,
  \href{https://doi.org/10.1016/0550-3213(86)90220-8}{\emph{Nucl. Phys. B}
  {\bfseries 278} (1986) 417}.

\bibitem{Sakata:1974hd}
I.~Sakata, \href{https://doi.org/10.1063/1.1666528}{\emph{J. Math. Phys.}
  {\bfseries 15} (1974) 1702}.

\bibitem{Dudek:2012gj}
J.~J. Dudek, R.~G. Edwards and C.~E. Thomas,
  \href{https://doi.org/10.1103/PhysRevD.86.034031}{\emph{Phys. Rev. D}
  {\bfseries 86} (2012) 034031}
  [\href{https://arxiv.org/abs/1203.6041}{{\ttfamily 1203.6041}}].

\bibitem{PhysRevD.69.054501}
{Morningstar, Colin and Peardon, Mike},
  \href{https://doi.org/{10.1103/PhysRevD.69.054501}}{\emph{{Phys. Rev. D}}
  {\bfseries {69}} ({2004}) {054501}}.

\bibitem{Follana:2006rc}
E.~Follana, Q.~Mason, C.~Davies et~al.,
  \href{https://doi.org/10.1103/PhysRevD.75.054502}{\emph{Phys. Rev. D}
  {\bfseries 75} (2007) 054502}
  [\href{https://arxiv.org/abs/hep-lat/0610092}{{\ttfamily hep-lat/0610092}}].

\bibitem{DeTar:2014gla}
C.~DeTar and S.-H. Lee,
  \href{https://doi.org/10.1103/PhysRevD.91.034504}{\emph{Phys. Rev. D}
  {\bfseries 91} (2015) 034504}
  [\href{https://arxiv.org/abs/1411.4676}{{\ttfamily 1411.4676}}].

\bibitem{Michael:1985ne}
C.~Michael, \href{https://doi.org/10.1016/0550-3213(85)90297-4}{\emph{Nucl.
  Phys. B} {\bfseries 259} (1985) 58}.

\bibitem{Luscher:1990ck}
M.~Luscher and U.~Wolff,
  \href{https://doi.org/10.1016/0550-3213(90)90540-T}{\emph{Nucl. Phys. B}
  {\bfseries 339} (1990) 222}.

\end{thebibliography}\endgroup
\end{document}